\documentclass[12pt]{iopart}
\usepackage{iopams}
\usepackage{graphicx}
\usepackage{epstopdf}

\newcommand{\ket}[1]{\left\vert{#1}\right\rangle}
\newcommand{\ketbra}[2]{\left| #1 \right\rangle\!\!\!\,\left\langle #2 \right|}

\newcommand{\expect}[1]{\left\langle{#1}\right\rangle}

\newcommand{\braket}[2] {\left \langle #1 \mid #2  \right \rangle}

\newcommand{\Hrk}{H_{\mathrm{RK}}}
\newcommand{\rttw}{\sqrt{12}\times\sqrt{12}}
\newcommand{\psicol}{$\vert \Psi_{col} \rangle$}
\newcommand{\psirttw}{$\vert \Psi_{\sqrt{12}} \rangle$}
\newcommand{\plaqa}{
 {\mathchoice
  {\includegraphics[height=1.6ex]{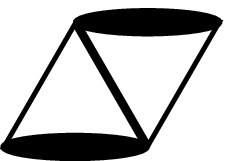}}
  {\includegraphics[height=1.6ex]{plaqa}}
  {\includegraphics[height=1.2ex]{plaqa}}
  {\includegraphics[height=0.9ex]{plaqa}}
 }
}
\newcommand{\plaqb}{
 {\mathchoice
  {\includegraphics[height=1.6ex]{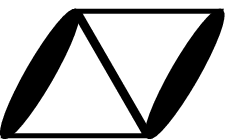}}
  {\includegraphics[height=1.6ex]{plaqb}}
  {\includegraphics[height=1.2ex]{plaqb}}
  {\includegraphics[height=0.9ex]{plaqb}}
 }
}

\begin{document}
\title[Loop condensation in the triangular lattice QDM]{Loop condensation in the triangular lattice quantum dimer model}

\author{C M Herdman$^{1}$ and K B Whaley$^2$}
\address{Berkeley Center for Quantum Information and Computation

Departments of Physics$^{1}$ and Chemistry$^{2}$

University of California, Berkeley, California 94720, USA}
\ead{herdman@berkeley.edu}

\date{\today}

\begin{abstract}
We study the mechanism of loop condensation in the quantum dimer model on the triangular lattice. The triangular lattice quantum dimer model displays a topologically ordered quantum liquid phase in addition to conventionally ordered phases with broken symmetry. In the context of systems with extended loop-like degrees of freedom, the formation of such topological order can be described in terms of loop condensation. Using Monte Carlo calculations with local and directed-loop updates, we compute geometric properties of the transition graph loop distributions of several triangular lattice quantum dimer wavefunctions that display dimer-liquid to dimer-crystal transitions and characterize these in terms of loop condensation.
\end{abstract}


\maketitle

\section{Introduction}
Quantum dimer models (QDMs) are among the simplest lattice models known to display exotic quantum phases~\cite{Rokhsar1988,Moessner2001c,Moessner2001d,Misguich2002,Moessner2008}. In addition to conventional, symmetry broken dimer crystal phases, QDMs have quantum liquid phases with no local order. In particular, the QDM on the triangular lattice has been shown to possess a quantum liquid ground state with topological order~\cite{Moessner2001d,Ralko2005a}. Unlike conventionally ordered phases, topological phases have no conventional broken symmetry and therefore can not be distinguished by a local order parameter~\cite{Wen1990,Nayak2008}. Features of topological order include a robust topological ground state degeneracy and a gap to anyonic quasiparticle excitations. Because of these unique features, topological phases have been proposed as the basis for a physically fault tolerant quantum computer~\cite{Freedman2002a,Kitaev2003,Nayak2008}. Several spin and Bose-Hubbard models are known to map to quantum dimer models~\cite{Balents2002a,Freedman2005a}; consequently there have been proposals to realize triangular lattice QDMs in various experimental systems as the basis for a topologically protected qubit~\cite{Misguich2005a,Ioffe2002,Albuquerque2008}.

A transparent physical picture of topological order exists for lattice models with loop degrees of freedom. Such quantum loop model models are known to possess topologically ordered phases~\cite{Kitaev2003,Levin2005a,Freedman2004,Fendley2005,Fendley2008}, and the formation of topological order can be described in terms of loop condensation. In the loop condensation picture, the creation of topological order coincides with the formation of fluctuating loops on all length scales. While the microscopic degrees of freedom in QDMs are non-overlapping dimers, QDMs can be mapped to closed loop models~\cite{Sutherland1988b,Kohmoto1988a}. In this paper we use the mapping of QDMs to closed loop models as a tool to analyze the phases of the triangular lattice QDM. In particular we consider the geometric properties of the loop distribution to understand the loop nature of the liquid and crystalline phases of the QDM.

The structure of this paper is the following: first we review the triangular lattice QDM and the concept of loop condensation and topological order in sections~\ref{sec:QDM} and~\ref{sec:loopcond}. We then introduce the loop mapping of the quantum dimer model in section~\ref{sec:dimloop}. The main results of this paper are then presented: we study several wavefunctions that are representative of phases of the triangular lattice QDM and capture transitions from the dimer liquid to dimer crystal phases. Using Monte Carlo sampling of these wavefunctions with local and directed-loop updates, we investigate the loop nature of these phases in sections~\ref{sec:RK}-\ref{sec:wilson}. This allows us to characterize the loop properties of the liquid and crystalline phases of the triangular lattice QDM.
 
\section{The quantum dimer model on the triangular lattice} 
\label{sec:QDM}

First we briefly review the quantum dimer model on the triangular lattice. In a hard-core dimer model, the degrees of freedom are dimers that live on the links of a lattice, and the hard-core constraint forbids more than a single dimer from touching each vertex. The quantum dimer model was first introduced on the square lattice by Rokhsar and Kivelson~\cite{Rokhsar1988}, and later generalized to the triangular lattice by Moessner and Sondhi~\cite{Moessner2001d}. The Hilbert space of the QDM comprises fully-packed, non-colliding dimer coverings of a lattice, where all dimerizations are by definition orthogonal. The only way to rearrange dimers in a fully-packed dimerization without violating the hard-core constraint is to flip dimers around a closed loop on the links of the lattice, where the links in the loop are alternately occupied and unoccupied by dimers. Consequently, the minimal quantum dynamics give a resonance between the two orientations of dimers around a "flippable" plaquette. On the triangular lattice a flippable plaquette is defined as a length $4$ rhombus lattice containing two parallel dimers. The canonical QDM Hamiltonian on the triangular lattice~\cite{Rokhsar1988,Moessner2001d} is:
\begin{equation}
\Hrk \equiv \sum_{p} -t \Bigl( \ketbra{\plaqa}{\plaqb} + h.c. \Bigr)+ v \Bigl( \ketbra{\plaqa}{\plaqa} +\ketbra{\plaqb}{\plaqb} \Bigr) \label{HRK}.
\end{equation}
In \eref{HRK}, the sum is over all rhombus plaquettes labeled by $p$, including all three orientations, and $\vert \plaqa \rangle$ and $\vert \plaqb \rangle$ represent the two possible flippable dimer configurations around $p$. The $t$ term gives the dimers kinetic energy, and the $v$ term represents an interaction between parallel dimers. In this work we will only consider $t>0$. 

These local dynamics are not ergodic over all dimer configurations, and consequently split the Hilbert space into topological sectors that are defined by quasi-ergodicity of the local dynamics. This means that on the triangular lattice there are a finite number of topological sectors, within which the plaquette flip dynamics are ergodic~\cite{Moessner2001d}. On a lattice with periodic boundary conditions, one can also classify topological sectors by the parity of the number of dimers crossing a closed loop that passes through the faces of the plaquettes and encircles the lattice; a plaquette flip conserves this parity. On a torus, there are 4 such parity sectors, two for each of the directions around the lattice. However, there also exist a finite number of symmetry related "staggered" configurations with no flippable plaquettes, which are frozen under plaquette flip dynamics; it is believed that plaquette flips are ergodic in each parity sector excluding these staggered configurations~\cite{Moessner2001d}. We label these parity winding sectors as $(E,E)$, $(E,O)$, $(O,E)$ and $(O,O)$, where $E$ and $O$ represent the even and odd parity sector, respectively.

At the point $v=t$, the so-called RK point, the exact zero energy ground state of $\Hrk$ is the equal superposition of all dimerizations:
\begin{equation}
\ket{RK} = \sum_C \ket{C} \label{RK}
\end{equation}
where the sum in equation \eref{RK} is taken over all dimer configurations $C$ in a topological sector~\cite{Rokhsar1988}. Since the norm of $\vert RK\rangle$ is equal to the partition function of the corresponding classical dimer model, expectation values of diagonal observables are identical to those of the classical dimer model. Dimer-dimer correlation functions are known to decay exponentially in the classical dimer model on the triangular lattice~\cite{Fendley2002,Ioselevich2002a}, and so $\vert RK \rangle$ describes a dimer liquid phase. Additionally, imaginary-time correlation functions of equation \eref{RK} can be related to dynamic correlation functions of a Monte Carlo simulation of the classical dimer model~\cite{Henley1997,Henley2004a,Castelnovo2005a}; on the triangular lattice, such calculations have shown that $\Hrk$ has a finite gap at the RK point~\cite{Ivanov2004}. Quantum Monte Carlo calculations by Moessner and Sondhi~\cite{Moessner2001d} and Ralko \textit{et. al.}~\cite{Ralko2005a} demonstrate that the dimer liquid phase extends beyond the RK point for a finite range of $v/t$ for $v < t$ and possesses $Z_2$ topological order. While the QDM has been related to $Z_2$ gauge theory in the literature, no exact mapping from the deconfined phase of a $Z_2$ gauge theory to the dimer liquid is known~\cite{Moessner2001,Misguich2008d}.

\begin{figure}[h] 
   \centering
  \includegraphics[width=3in]{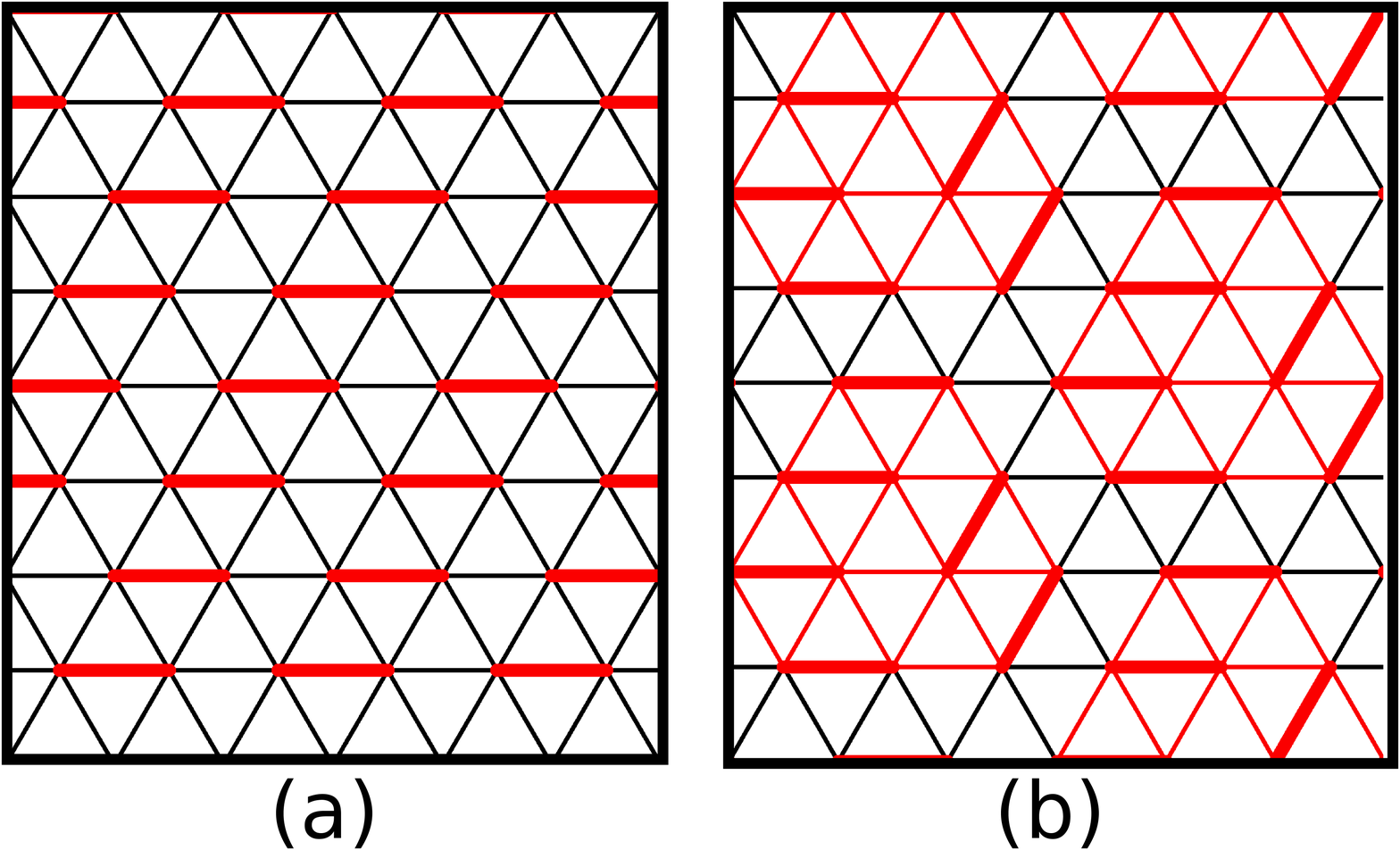} 
 \caption{ Reference dimer configurations that are representative of dimer crystals in the triangular lattice QDM. The thick red dimers represent the reference configurations, $R_0$. (a) A columnar configuration that defines $R_{col}$. (b) In the $\rttw$ phase dimers resonate in 12 site hexagons, colored red in this figure. The thick red dimers show the reference configuration $R_{\sqrt{12}}$.
}\label{fig_rt12_col_lats}
\end{figure}

For $v>t$, the triangular lattice QDM is in a staggered dimer crystal phase, as the staggered configurations are frozen, zero energy ground states of $\Hrk$. For $\vert v \vert >> t$ and $v/t < 0$, states with the maximum number of flippable plaquettes are favored. The 12 symmetry related columnar configurations (see figure \ref{fig_rt12_col_lats}) have the maximum number of flippable plaquettes, $N_l/6$, where $N_l$ is the number of links in the lattice. Additionally, any configuration reached by translating dimers along any number of rows of a columnar configuration and by rotating all dimers in any number of columns of a columnar configuration, generates another maximally flippable dimerization\cite{Moessner2001d}. References \cite{Moessner2001d} and \cite{Ralko2005a} show that quantum fluctuations favor the columnar order, and that for $v/t \lesssim -0.75$, $\Hrk$ has a columnar dimer crystalline ground state. The staggered and columnar phases break both translational and rotational symmetries of the lattice. 

Between the columnar crystal and the dimer liquid lies a resonating dimer crystal phase with a 12-site unit cell (see figure \ref{fig_rt12_col_lats}) termed the $\rttw$ phase\cite{Moessner2001d,Ralko2006,Ralko2005a}. In this phase, dimers resonate within 24 link hexagons, and most translational symmetries are broken. However this phase retains a rotational symmetry of the lattice. Ralko et. al. showed that the  dimer correlation functions are qualitatively reproduced by a wavefunction that is an equal superposition of all dimerizations within the hexagons~\cite{Ralko2006,Misguich2008d}.

\section{Loop condensation}
\label{sec:loopcond} 

Here we briefly review the paradigm of loop condensation in topological phases. Loop condensation has been discussed as a physical mechanism for the formation of topological order by several authors~\cite{Levin2005a,Kitaev2003,Fendley2008,Freedman2004}. Quantum systems with loop-like degrees of freedom may display a variety of phases. These include symmetry broken loop crystal phases for which the length of loops is peaked at particular length scale, as well as scale-invariant loop liquid phases characterized by fluctuating loops on all length scales.

One of the simplest exactly solvable models with a quantum liquid ground state is the toric code \cite{Kitaev2003}:
\begin{eqnarray}
H_{\mathrm{TC}} &= -\lambda_e \sum_v A_v - \lambda_m \sum_p B_p  \label{H_TC}, \qquad A_v \equiv \prod_{j \in v} \sigma^z_j, \quad B_p \equiv  \prod_{j\in p} \sigma^x_j.
\end{eqnarray}
Here, $\{\sigma_j\}$ are spin-$1/2$ degrees of freedom that are located on the links of a square lattice on a torus. $v$ and $p$ label the vertices and plaquettes of the lattice. When $\lambda_e = \lambda_m$, the ground state, $\vert \Psi_{\mathrm{TC}} \rangle$, is an eigenstate of all $A_v$, i.e. $A_v\ket{\Psi_{\mathrm{TC}}}=+\ket{\Psi_{\mathrm{TC}}}$ for all $v$. These eigenstates of the vertex operators $A_v$ have 0, 2 or 4 down spins touching each vertex. The ground state subspace can be interpreted in terms of a closed loop model by choosing a reference configuration $\vert R \rangle$ in the $\sigma^z$ basis (from the subspace of $+1$ eigenstates of all $A_v$) that is defined as the empty loop state. For example, we may choose $\vert R \rangle$ to be the spin polarized state, $\ket{\{\forall \sigma_j^z = +1\}}$. Any other configuration in the $A_v = +1$ subspace can be reached by applying a product of closed loop operators:
\begin{eqnarray}
\ket{A}  &= \prod_{ \ell \in \left \{ \ell_A \right \}_R } W_{\ell} \ket{R}, \quad W_{\ell} \equiv \prod_{j \in \ell} \sigma^x_j \label{TCloops},
\end{eqnarray}
where $\ell$ is a closed loop along the links and $ \left \{ \ell_A \right \}_R$ a closed loop covering of the square lattice. Since $\left[ A_v, W_{\ell} \right] = 0, \forall v$, $\vert A \rangle$ is in the $A_v = +1$ subspace for all $\left\{ \ell_A \right\}$. In this way, the closed loop covering defining $\vert A \rangle$ is $\left\{ \ell_A \right\}_R$. 
 
 In general $ \left \{ \ell_A \right \}_R$ may comprise both contractible loops and non-contractible loops that wind around the torus. The winding sector $(w_1,w_2)_A$ is defined by the number of windings in $ \left \{ \ell_A \right \}_R$ about the two axes of the torus. Though the the plaquette term $B_p$ causes fluctuations in these loop coverings, the parity of the winding number is conserved. The operator
 \begin{equation}
 \tilde{W}_{1,2} \equiv \prod_{j \in \tilde{c}_{1,2}} \sigma^z_j
 \end{equation}
 measures the parity of the winding sector and commutes with $H_\mathrm{TC}$, where $\tilde{c}_{1,2}$ are loops that pass through the faces of the plaquettes and wind about one of the axes of the torus. Therefore, the ground state subspace is divided into four topological sectors that are defined by the parity of these two winding numbers.
 
The ground state $\vert \Psi_{\mathrm{TC}} \rangle$ is an equal superposition of all loop coverings in a given topological sector, and has fluctuating loops on all length scales. This quantum loop gas has no local order, but does possess topological order and is described by the deconfined phase of $Z_2$ gauge theory~\cite{Kitaev2003}. By considering the limit $\lambda_e \rightarrow \infty$, perturbations can drive transitions within the closed loop subspace \cite{Trebst2007}. For example addition of a magnetic field $H' = -h \sum_j \sigma^z_j$ will drive a transition to a spin polarized phase: for $h > \left \vert h_c \right \vert$ the ground state is a dilute loop crystal, and for $h < -\left \vert h_c \right \vert$ the ground state is a fully packed loop crystal. These transitions do not involve spontaneous symmetry breaking; we can alternately consider the effect of adding a loop interaction such as $H' = J \sum_{\left \langle i,j \right \rangle } \sigma^z_i \sigma^z_j$ which favors a rotational symmetry broken loop crystal for $J \gg \lambda_m$. We note that loop crystals need neither be dilute nor involve short loops; however, we may choose $\vert R \rangle$ to reflect the broken symmetries of the crystal phase, such that the resulting loop configuration will involve only short loops.

\section{Loop order in quantum dimer models}
\label{sec:dimloop} 

\begin{figure}[h] 
   \centering
  \includegraphics[width=2.5in]{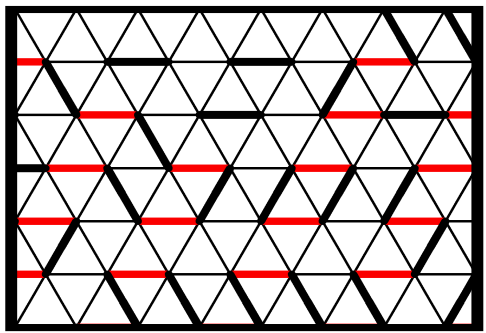} 
 \caption{ A transtion graph loop configuration $\{C:R_0\}$ formed by superimposing a physical dimer configuration $\ket{C}$ (black) with the reference dimerization $\vert R_{col} \rangle$ (red).
}\label{fig_col_transloop}
\end{figure}

A dimerization of a lattice, $\ket{C}$, can be interpreted as a closed loop covering of the lattice by superimposing it with an arbitrary reference dimerization, $\vert R_0 \rangle$ (see figure \ref{fig_col_transloop})~\cite{Sutherland1988b,Kohmoto1988a}. The resulting doubled dimerization, $\{C:{R_0}\}$, known as the transition graph, has both a reference and physical dimer touching each vertex; this therefore forms closed loops on the lattice with the exception of where dimers in $\ket{C}$ coincide with those of $\vert R_0 \rangle$. It is necessary to choose a convention for the definition of such coinciding dimers. Two possible interpretations of overlapping dimers are: (a) they are considered to be a loops of length 2, or (b) they are defined as an empty links of the transition graph. Choosing option (a) means that the transition graph is a fully packed loop covering of the lattice and the transition graph $\{R_0:{R_0}\}$ comprises all length 2 loops. Choosing (b) means that the transition graph is not fully packed, and that $\{R_0:{R_0}\}$ is an empty loop covering. In order to make an analogy with the definitions of loops given in equation \eref{TCloops}, we will generally choose below the convention (b), except where noted.

Choosing a reference dimerization $\vert R_0 \rangle$ translates a dimer configuration into a loop configuration. However, unlike the toric code, there is no reference configuration that retains all of the lattices symmetries. Therefore we are forced to choose an $\vert R_0 \rangle$ that breaks translational and rotational symmetries of the lattice. Certain properties of the resulting loop model will depend on this choice of a symmetry broken $\vert R_0 \rangle$. All other dimerizations can be reached from $\vert R_0 \rangle$ by applying a product of closed loop operators that alternately pass through the links that are occupied and unoccupied by dimers in $\vert R_0 \rangle$:
\begin{eqnarray}
\ket{C} = \prod_{\ell \in \left\{ \ell_C\right\}_{R_0}} W_{\ell} \ket{R_0},\quad W_{\ell} \equiv \prod_{ \left( r,s \right) \in \ell} d^-_r  d^+_s. \label{Wtrans}
\end{eqnarray}
In equation \eref{Wtrans}, $d^+_j$($d^-_j$) are dimer creation (annihilation) operators and $(r,s)$ are a pair of links in $\ell$ that meet at a vertex, that are occupied and unoccupied, respectively, by a dimer in $\vert R_0 \rangle$. The set $\{ \ell_C \}_{R_0}$ is simply the transition graph $\{C:R_0\}$, where we interpret links that are occupied in both in $\ket{C}$ and $\vert R_0 \rangle$ as empty links in $\{ \ell_{R_0} \}$. We note that the choice of $\vert R_0 \rangle$ restricts the possible loop coverings such that if a loop touches a vertex, it must pass through the link occupied in $\vert R_0 \rangle$. This dimer-loop Hilbert space is not equivalent to the Hilbert space of all loop coverings of the lattice. 

The phases of $\Hrk$ may now be re-interpreted as those of a loop model. $\vert RK \rangle$ is the equal superposition of all loop coverings that pass through the dimers of $\vert R_0 \rangle$, and therefore is a loop gas. We expect the topological phase adjacent to the RK point to be described by a scale-invariant loop liquid. Static dimer crystal phases, such as the columnar and staggered phases, correspond to loop crystals with a narrow distribution of loop lengths; if $\vert R_0 \rangle$ is chosen to be a dimer configuration with the corresponding broken symmetry, then the loop model is in a short looped dilute phase. The loop order in the resonating $\rttw$ phase is less obvious: if $\vert R_0 \rangle$ is chosen to be one of the dimerizations with $\rttw$ order, the ideal $\rttw$ state will be a dense, fluctuating short looped phase. This transition loop description of the dimer model allows us to make a connection to the loop condensation picture of topological order. A dimer crystal to dimer liquid transition is then described as a transition from a loop crystal to a scale-invariant loop liquid.

\section{The dimer liquid at the RK point} 
\label{sec:RK}

Here we analyze the geometric properties of the loop description of the RK wavefunction, taking this to be representative of the dimer liquid. This analysis follows a similar approach as that of Sutherland and Kohmoto \& Shapir in their studies of the spin-$1/2$ RVB wavefunction on the square lattice~\cite{Sutherland1988b,Kohmoto1988a,Kohmoto1988,Sutherland1988,Sutherland1988a,Shapir1989}. To understand the loop properties of $\vert RK \rangle$, we can refer to the classical $O(n)$ loop model~\cite{Domany1981,Nienhuis,Nienhuis1984}. In this model, which can be solved exactly on the honeycomb lattice, configurations are closed loop coverings of the lattice, and the partition function is given by
\begin{equation}
Z_{O(n)} = \sum_C K^{\mathcal{L}}n^{\mathcal{N}}, \label{ZOn}
\end{equation}
where $C$ is a closed loop configuration comprising $\mathcal{N}$ loops of total length $\mathcal{L}$. In equation \eref{ZOn}~, $K$ is the weight per length of loops and $n$ is the loop fugacity. For $n \leq 2$, there are two phases separated by a critical line at $K_c = [2+\sqrt{2-n}]^{-1/2}$. $K_c$ separates a dense loop phase for $K > K_c$ and a dilute phase for $K < K_c$. The dense phase is characterized by a power law distribution of loop lengths, $P_\ell \left(s\right) \sim s^{-p}$, where $P_\ell \left(s\right)$ is the density of loops of length $s$~\cite{Ding2007}. Additionally, there is a large spanning loop $\ell_M$ with fractal dimension $D_f$, whose length scales with the system size $L$ as $s(\ell_M) \sim L^{D_f}$~\cite{Saleur1987}. These exponents are related by the scaling dimension $x$: 
\begin{eqnarray}
D_f = 2-x,\qquad p = 1+\frac{2}{2-x}
\end{eqnarray}
Within the dense phase, $x$ is given by
\begin{eqnarray}
x = \frac{g}{2}-\frac{1}{2g}\left(g-1\right)^2, \quad n = -2 \mathrm{\cos} \left( \pi g \right)
\end{eqnarray}
for $0 < g \leq 1$ where $g$ is the coupling constant of the Coulomb gas description of of the $O(n)$ model~\cite{Nienhuis1984}. The $O(n)$ loop model on the triangular lattice has been studied by Knops \textit{et. al.}, and is known to possess a critical dense phase of the same universality class as that of the honeycomb lattice~\cite{Knops1998}.

In the $O(1)$ model on the honeycomb lattice, $K_c = 1/\sqrt{3}$, so $K=1$ is in the dense phase. At $K=1$, $Z_{O(1)}$ is an equal weighted sum over all loop configurations with the critical exponents $D_f = 7/4$ and $p = 15/7$~\cite{Nienhuis1984,Saleur1987}. The RK wavefunction can similarly be described as equal superposition of all transition loop configurations within a topological sector, for a given choice of $\vert R_0 \rangle$, and therefore we can relate the loop configuration at the RK point to the critical phase of $O(1)$ model.

To calculate the loop properties of $\vert RK \rangle$ we choose a reference configuration $\vert R_0 \rangle$, and then compute the distribution of loops in the transition graph, $P^{R_0}_\ell(C,s) = h_{\ell}^{R_0} (C,s) / L^2$ where $h_{\ell}^{R_0} (C,s)$ is the number of loops of length $s$ in $\{C:R_0\}$ and $L$ is the linear dimension of the lattice. We have computed the expectation value of $P_{\ell}^{R_0}$ by Monte Carlo sampling of $\vert RK \rangle$ using local plaquette flip updates on triangular lattices of up to $L=192$, with $N_l = 3*L^2$ links.
\begin{figure}[h] 
   \centering
  \includegraphics[width=3.5in]{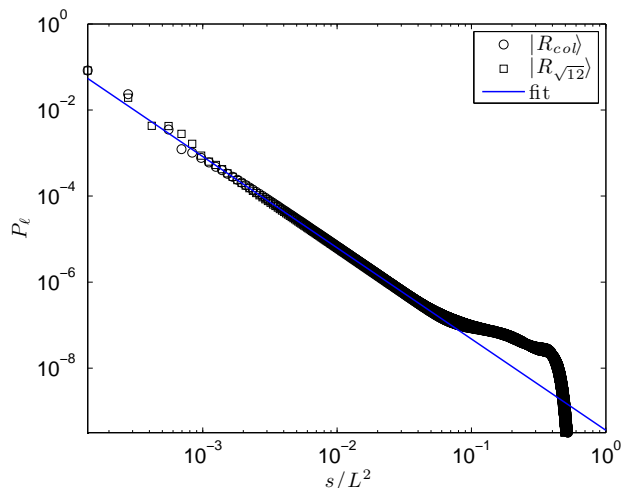} 
 \caption{ 
Loop distribution $P_{\ell}$ vs. loop length $s$ at the RK point on a $L=120$ triangular lattice using reference configurations $\vert R_{col} \rangle$ and $\vert R_{\sqrt{12}} \rangle$ in the $(E,E)$ winding sector. The line shows the fit to a power law decay.
}\label{figRK_Pl}
\end{figure}

\begin{figure}[]
\begin{center}
$\begin{array}{c c}
\includegraphics[width=3in]{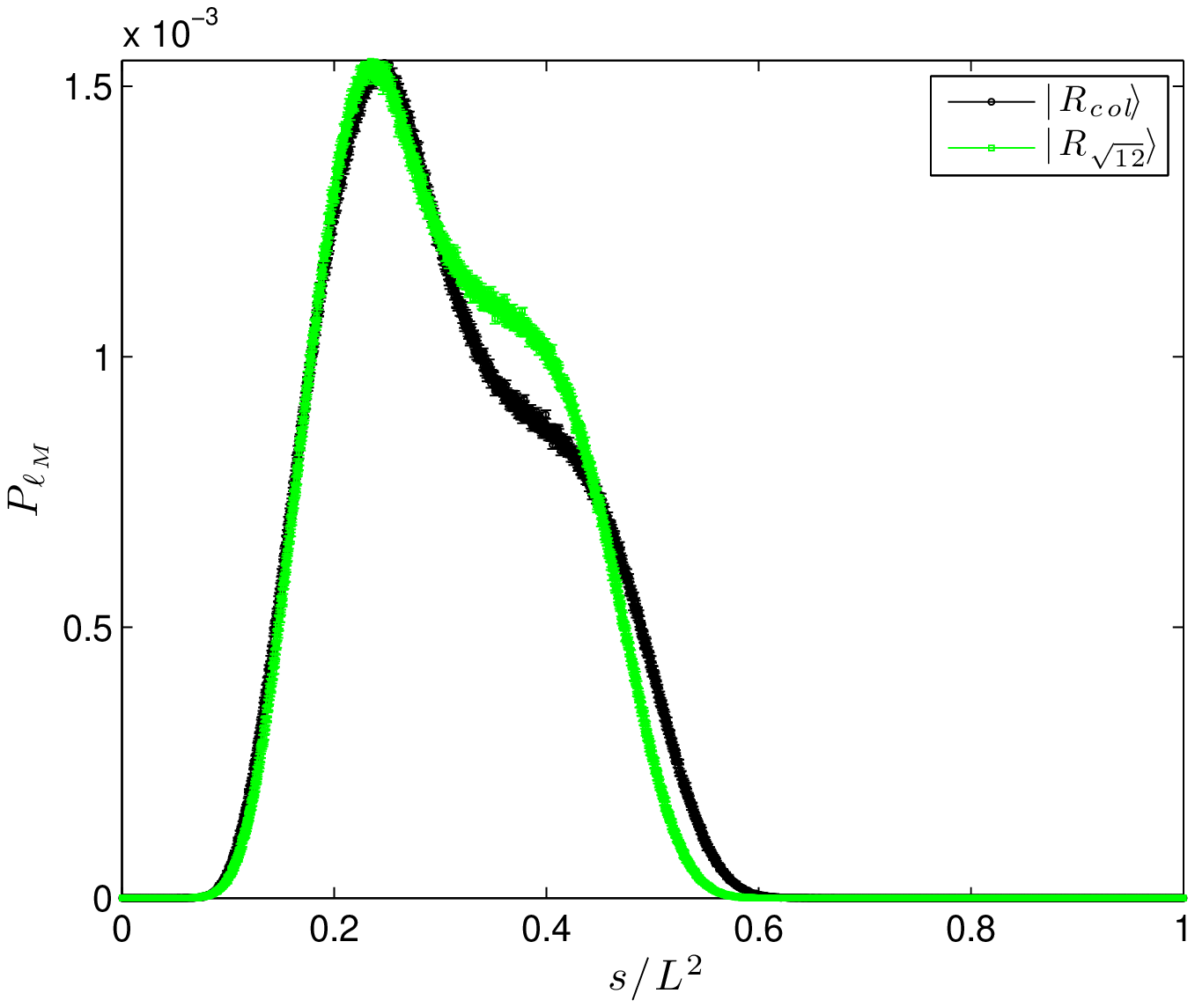} &
\includegraphics[width=3in]{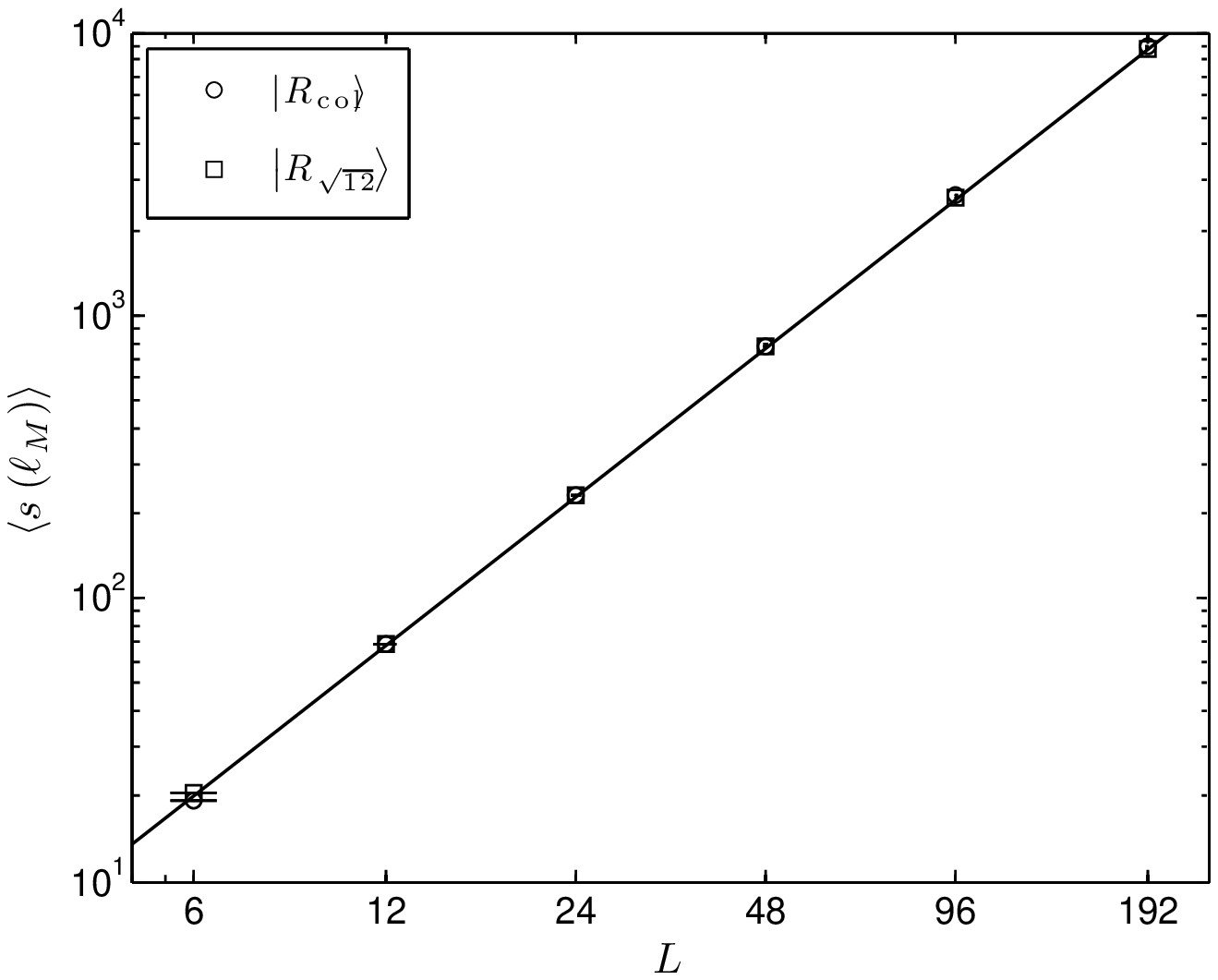} \\ 
\mbox{\bf (a)} & \mbox{\bf (b)}
 \end{array}$
 \end{center}
  \caption{ (a) Longest loop distribution, $P_{\ell_M}$ vs. loop length $s(\ell_M)$ at the RK point using reference dimerizations $\vert R_{col} \rangle$ and $\vert R_{\sqrt{12}} \rangle$, in the $(E,E)$ winding sector on lattice size $L=72$. (b) Finite size scaling of the expectation value of the length of the longest loop $s(\ell_M)$ at the RK point for $\vert R_{col} \rangle$ and $\vert R_{\sqrt{12}} \rangle$, in the $(E,E)$ winding sector of lattice sizes up to $L = 192$. Errorbars are smaller than the symbols shown. The line shows the fit to $s(\ell_M) \propto L^{D_f}$. 
}\label{figRK_lM}
\end{figure}

Here we will consider two choices of reference configurations: $\vert R_{col} \rangle$, the columnar reference state shown by the red links in figure \ref{fig_rt12_col_lats}a, and $\vert R_{\sqrt{12}} \rangle$, which is the configuration illustrated by the thick red links in figure \ref{fig_rt12_col_lats}b that is one of the equally weighted configurations of the ideal $\rttw$ crystal. In figure \ref{figRK_Pl}, the loop distribution $P_{\ell}(s)$ is plotted for these choices of $\vert R_0 \rangle$. We see a clear power law over a range of length scales. The best fit power law $p = 2.12 \pm 0.01$; this is consistent with the theoretical value for the $O(1)$ model, $p = 15/7\simeq2.14$~\cite{Nienhuis1984,Ding2007}, considering the imperfect power law behavoir of $P_\ell$ on lattices of these sizes. Additionally, we see that $P_{\ell}(s)$ is independent of the choice of $\vert R_0 \rangle$ for lengths longer than about $16$. We have computed the distribution of the longest loop $\ell_M$, $P_{\ell_M} (s)$, and this is plotted in figure \ref{figRK_lM}a. To compute $D_f$ at the RK point, we analyze the finite size scaling of $\ell_M$, shown in figure \ref{figRK_lM} (b). The best fit gives $D_f = 1.751\pm.001$, which agrees with the known fractal dimension of the $O(1)$ model, $D_f = 7/4$~\cite{Nienhuis1984,Saleur1987}.

We can also characterize the loop liquid in terms of the total loop density, $\rho = \mathcal{L} / \mathcal{L}_M$, where $\mathcal{L}_M = L^2$ is the maximum possible total loop length. The value $\rho = 1$ corresponds to the maximum loop density, where no dimers reside on links occupied in $\vert R_0 \rangle$, while $\rho$ strictly vanishes only for $\vert R_0 \rangle$. Consequently, in crystalline phases, $\rho$ depends on the choice of $\vert R_0 \rangle$. At the RK point, all links are occupied with probability $1/6$, and the total loop length is twice the average number of dimers on links unoccupied in $\vert R_0 \rangle$, which is $5/6 \times N_l/6$. Therefore at the RK point, $\rho = 5/6$; this agrees with our numerically computed value of $\rho = 0.83333(3)$ on a $L =120$ lattice, computed for $\vert R_{col} \rangle$ and $\vert R_{\sqrt{12}} \rangle$. The RK wavefunction can be described as a dense loop gas. We note that in this loop gas phase, both the total density and geometric properties of loops are independent of the choice of $\vert R_0 \rangle$.

\section{Loop order in dimer crystal phases}
\label{sec:crysloop}

Here will consider the loop order in transitions from the dimer liquid at the RK point to two dimer crystals that are present in phase diagram of the cannonical QDM as defined in equation \eref{HRK}. To analyze these transitions, we will consider a wavefunction that interpolates between $\vert RK \rangle$ and a given dimer crystalline order: 
\begin{equation}
\ket{\Psi \left(z\right)}  = \frac{1}{\sqrt{Z}} \sum_C z^{N_B} \ket{C} \label{psi_z}
\end{equation}
Here we will consider two possible colorings of the triangular lattice where the red links define links that are favored by the crystalline order, to which we give weight $1$, and $N_B$ is the number of occupied black links, which are given a weight $z$. In equation \eref{psi_z}, $Z\equiv \sum_C z^{2N_B \left(C \right)}$, which is the partition function of a non-interacting classical dimer model where black links are given a fugacity $z^2$. Additionally, $\ket{\Psi \left(z\right)}$ is the zero energy ground state of a local, stochastic matrix form~\cite{Ardonne2004,Castelnovo2005a} Hamiltonian that is generalization of $\Hrk$:
\begin{eqnarray}
H_z \equiv \sum_{p} -&t \Bigl( \ketbra{\plaqa}{\plaqb} + h.c. \Bigr)+t \Bigl( z^{\Delta N_B} \ketbra{\plaqa}{\plaqa}+ z^{-\Delta N_B}\ketbra{\plaqb}{\plaqb} \Bigr) \label{Hlam}
\end{eqnarray}
where $\Delta N_B \equiv N_B(\plaqb) - N_B(\plaqa)$. For $z = 1$, $\ket{\Psi \left(z = 1\right)}$ is the RK wavefunction, and and for $z = 0$, $\ket{\Psi \left(z = 0\right)}$ is an ideal dimer crystal as defined by the red links of the lattice.  

As described in section \ref{sec:QDM}, $\Hrk$ displays two crystalline phases for $v < t$, in addition to the dimer liquid phase. These crystal orders are illustrated in figure \ref{fig_rt12_col_lats}: the red links of panel (a) show one of 12 symmetry related columnar states, and the red links of panel (b) are favored by the $\rttw$ resonating crystalline order. We will use these two colorings to define two wavefunctions described by equation \ref{psi_z}: \psicol and \psirttw. We will use $\vert R_{col} \rangle$ and $\vert R_{\sqrt{12}} \rangle$ respectively as the reference configurations to define the transition loops. We have computed the loop distributions for both from Monte Carlo sampling of these wavefunctions.
 
 We can relate $\ket{\Psi(z)}$ to the $O(1)$ model by considering \psicol. In this case, since only the occupied links in $R_{col}$ are colored red, the total loop length $\mathcal{L} = 2*N_B$, and $-Log(z)$ acts as an effective loop tension. Accordingly, we can relate $z$ to $K$ in the $O(1)$ model. As discussed in section \ref{sec:RK}, $z=1$ sits in the dense phase, corresponding to the dimer liquid, and for some $z_c<1$ we expect to cross a phase transition into the dilute loop phase, corresponding to the dimer crystal. The relationship is less direct for \psirttw, where loop segments within the red hexagons are given unity weight, and only loop segments passing through black links are given a weight $z < 1$. However we still expect a transition from the long loop dense phase, to a short loop phase of finite loop density.
 
\begin{figure}[] 
   \centering
  \includegraphics[width=5.5in]{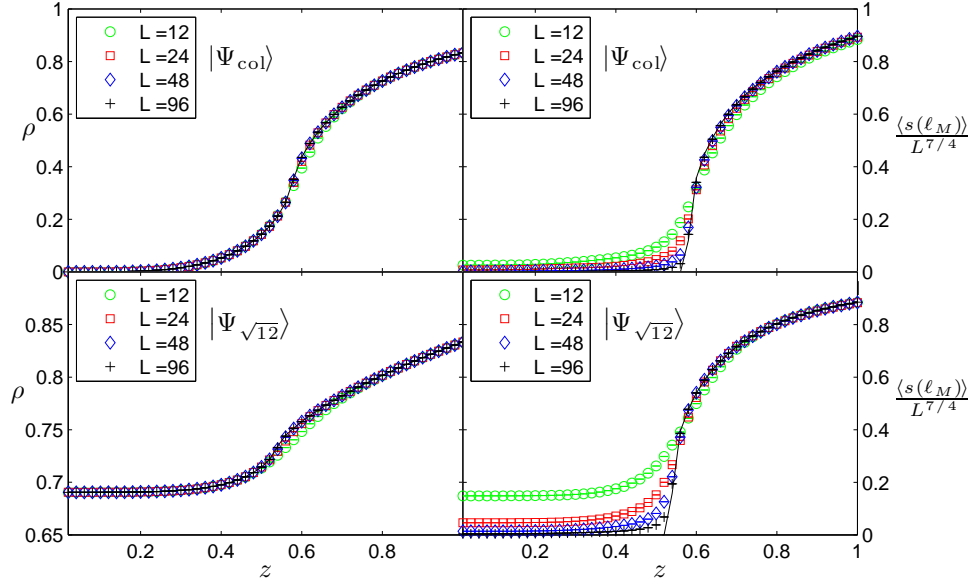} 
 \caption{ Top two panels: The total loop density $\rho$, and and longest loop length $s(\ell_M)$ for \psicol. Bottom two panels: The total loop density $\rho$, and and longest loop length $s(\ell_M)$ for \psirttw. The solid lines are extrapolations to the thermodynamic limit. We have rescaled $s(\ell_M)$ by $L^{7/4}$ for data collapse in the liquid phase.
}\label{fig_colrt12_rholmax_vp}
\end{figure}

\begin{figure}[]
\begin{centering}
$\begin{array}{c c}
  \includegraphics[width=3in]{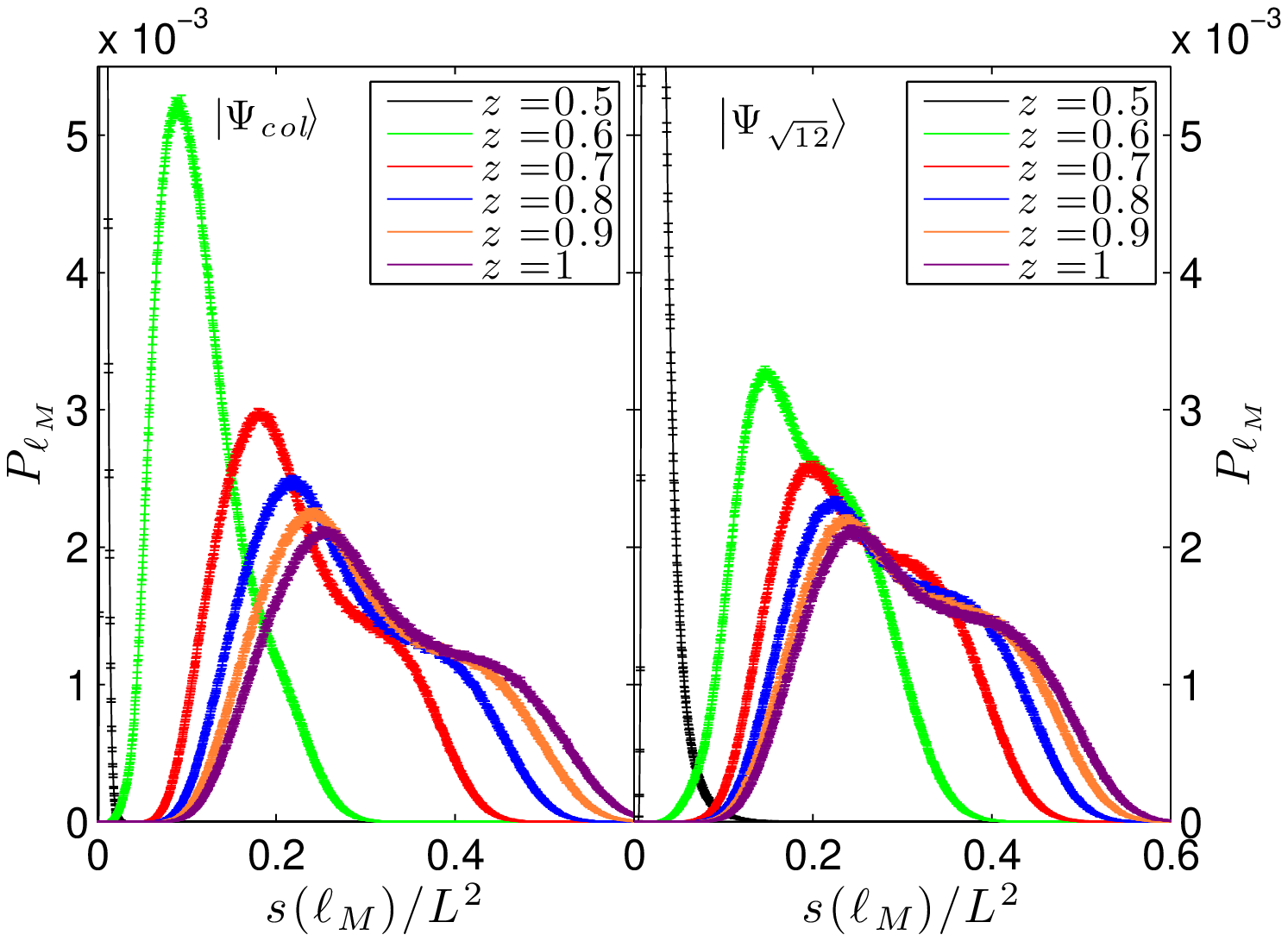} &
  \includegraphics[width=3in]{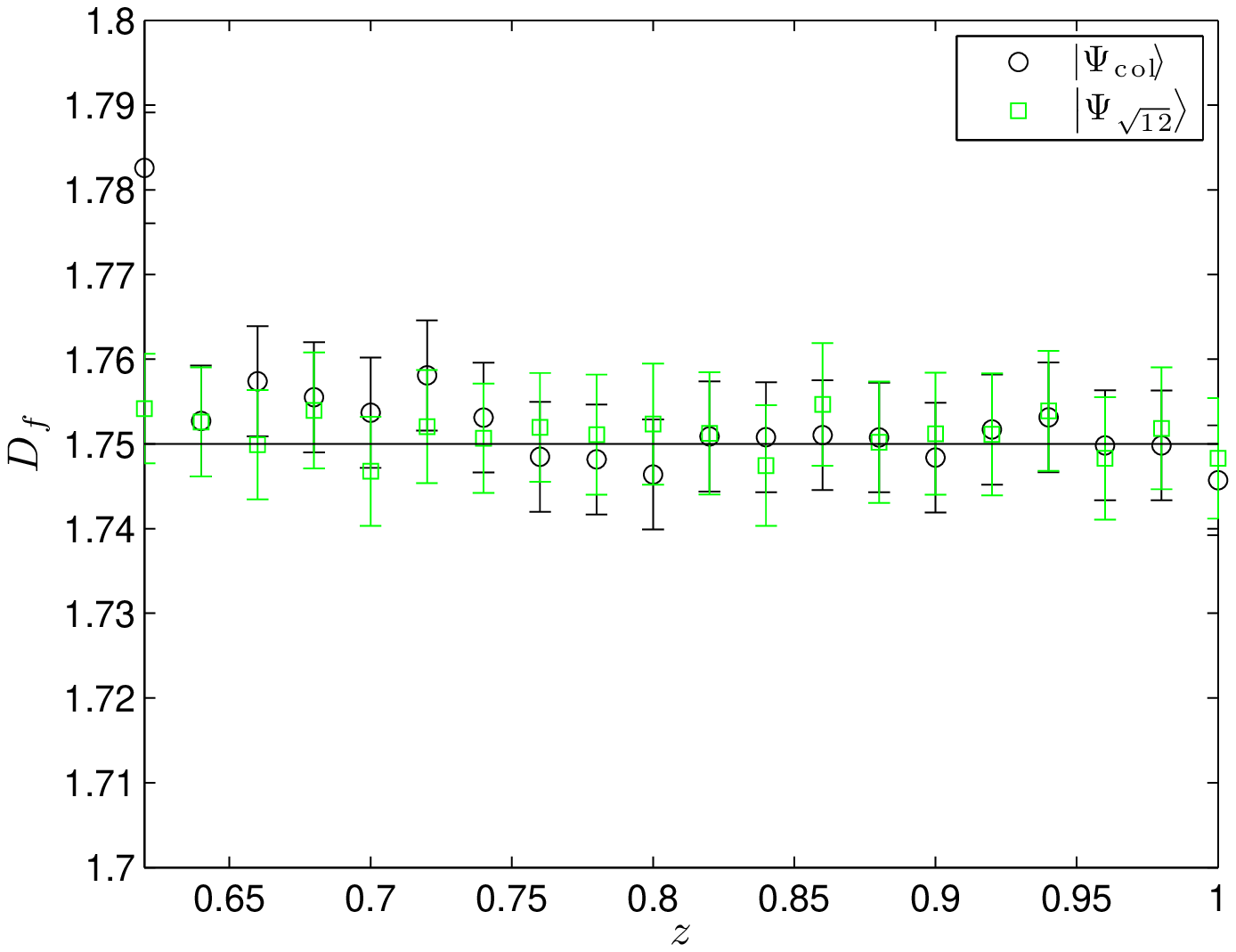} \\
  \mbox{\bf (a)} & \mbox{\bf (b)}
\end{array}$
\end{centering}
\caption{
(a) Longest loop distribution $P_{\ell_M}$ of \psicol~(left) and \psirttw~(right) plotted for several values of $z$ for a $L = 60$ lattice. (b) Fractal dimension $D_f$ of \psicol~and \psirttw~in the dimer liquid phase computed from the finite size scaling of $s(\ell_M)$ with lattice sizes up to $L = 144$. The line shows the value at the RK point, $D_f = 7/4$. The accuracy of $D_f$ is limited by the lattice sizes studied.
}\label{fig_colrt12_lM}
\end{figure}

In figure \ref{fig_colrt12_rholmax_vp}, the left panels show the total loop density $\rho$, and the right panels show the expectation value of the longest loop length, $s(\ell_M)$ as a function $z$, for \psicol (top) and \psirttw (bottom). Here we have rescaled $s(\ell_M)$ by $L^{7/4}$ where $7/4$ is the fractal dimension of the dimer liquid. In both cases $s(\ell_M)/L^{7/4}$ remains finite for $z$ close to $1$ and vanishes for small $z$. Finite size scaling of $s(\ell_M)$ suggests that the corresponding phase transitions occur at $z_c^{col} \approx 0.57$ for \psicol~and $z_c^{\sqrt{12}} \approx 0.53$ for \psirttw. For \psicol, $\rho$ vanishes in the columnar phase, whereas $\rho$ remains finite in the $\rttw$ phase of \psirttw. This distinction is due to the fact that in the columnar phase, dimers are pinned to $\vert R_{col} \rangle$, whereas the $\rttw$ phase dimers resonate within the hexagons and are not pinned to $\vert R_{\sqrt{12}} \rangle$. The location of these transitions may be compared with the transition of the honeycomb lattice O(1) model, where $K_c = 1/\sqrt{3}\simeq 0.58$. Figure \ref{fig_colrt12_lM}a shows the longest loop distributions $P_{\ell_M}(s)$ for \psicol~(left) and \psirttw~(right) for several values of $z$. We see that in both cases the broad distribution, which is a signature of the liquid phase, vanishes below the transition at $z_c$. The fractal dimension within the dimer liquid phase is plotted in figure \ref{fig_colrt12_lM}b. Deviations from $D_f = 7/4$ in the liquid phase near the transition may be due to the limitations of extrapolating  the finite size scaling on lattices of these sizes.

\begin{figure}[] 
   \centering
  \includegraphics[width=3in]{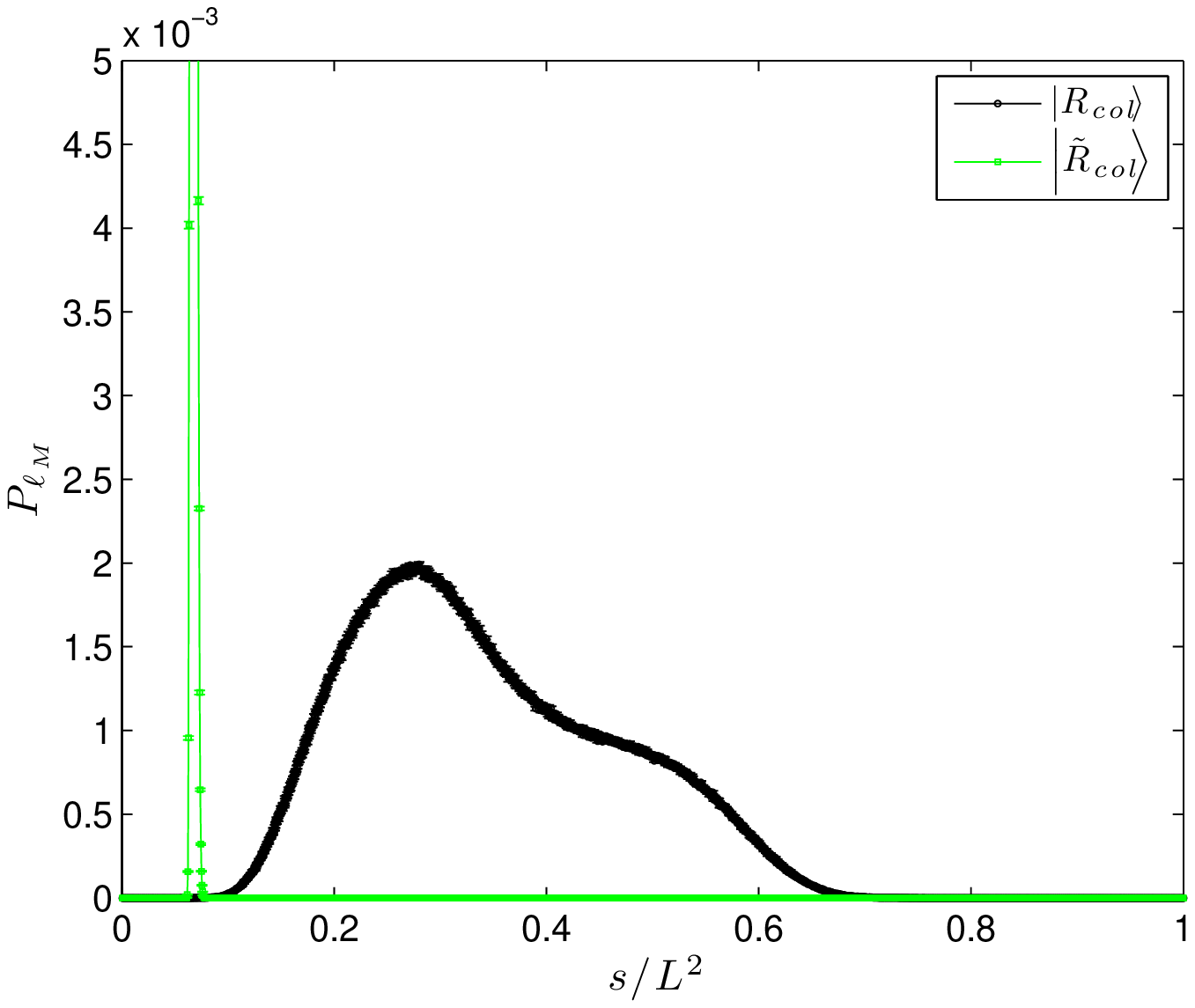} 
 \caption{ The longest loop distribution $P_{\ell_M}$ in the ideal $\rttw$ crystal, $\vert \Psi_{\sqrt{12}} (z=0) \rangle$ computed for two different columnar reference configurations $\vert R_{col} \rangle$ and $\vert \tilde{R}_{col} \rangle$. In a resonating crystal phase, for some choices of $\vert R_0 \rangle$, $P_{\ell_M}$ may clearly indicate crystalline order (seen by the peak at a single length scale for the choice $\vert \tilde{R}_{col} \rangle$) whereas for other choices (e.g. $\vert R_{col} \rangle$), the distribution $P_{\ell_M}$ may appear to be that of a liquid.
}\label{fig_rt12col_PlM}
\end{figure}

We note that in the crystalline phases, the loop distributions depend on the choice of $\vert R_0 \rangle$; here we have chosen reference dimerizations which reflect the known ordering in each phase. In particular, the vanishing of $s(\ell_M)/L^{D_f}$ in the crystalline phases is due to a choice of $\vert R_0 \rangle$ that is commensurate with the order. While in static crystals such as the columnar phase we may expect that any choice of $\vert R_0 \rangle$ will lead to $s(\ell_M)$ scaling as $L^0$, $L^1$, or $L^2$, in a resonating dimer crystal such as $\rttw$, this is not in general true. In fact, we find that for an arbitrary choice of $\vert R_0 \rangle$ the loop distribution in a resonating crystal phase may appear to be that of a liquid phase. This is illustrated in figure \ref{fig_rt12col_PlM} where we have computed $P_{\ell_M}$ for the ideal $\rttw$ state $\ket{\Psi_{\sqrt{12}}(z=0)}$ for two different columnar reference configurations: $\vert R_{col} \rangle$ (seen in figure \ref{fig_rt12_col_lats}a) and $\vert \tilde{R}_{col} \rangle$ which is related to $\vert R_{col} \rangle$ by translating all dimers one link horizontally. Figure \ref{fig_rt12col_PlM} shows that for $\vert R_{col} \rangle$, $P_{\ell_M}$ appears to be a liquid phase, but for$\vert \tilde{R}_{col} \rangle$ it is sharply peaked at a single length scale. In fact finite size scaling of the length of $\ell_M$ shows that for $\vert R_{col} \rangle$, the fractal dimension is that of the liquid phase ($D_f = 7/4$), but for $\vert \tilde{R}_{col} \rangle$ $D_f = 1$, indicative of the crystal order. This suggests that one may distinguish a liquid phase by computing $D_f$ for all (in this case 12) configurations that are related by symmetry to $\vert R_0 \rangle$, to ensure that a resonating crystalline order is not hidden by the choice of $\vert R_0 \rangle$. However, for any symmetry broken state, if $\vert R_0 \rangle$ is chosen to be commensurate with the broken symmetry, the length of $\ell_M$ will be finite.

\section{Interacting dimer loop liquid}
\label{sec:intdimer}

Section \ref{sec:crysloop} discusses the addition of an effective loop tension to the RK dimer liquid. Here we consider the effects of a translationally invariant dimer interaction by defining a generalization of the RK wavefunction:
\begin{eqnarray}
\ket{\Phi_{\alpha}} = \frac{1}{\sqrt{Z}} \sum_{C} \alpha^{-N_f \left(C \right) } \ket{C} \label{Phi_col}
\end{eqnarray}
In (\ref{Phi_col}), $N_f \left(C \right)$ is the number of flippable plaquettes in $\ket{C}$. For $\alpha = 1$,  $\ket{\Phi_{\alpha = 1}} \equiv \ket{RK}$; for $\alpha \ll 1$, configurations with the maximum number of plaquettes are favored. The square norm of $\ket{\Phi_{\alpha}}$ is the partition function of an interacting classical dimer model \cite{Trousselet2007}:
\begin{eqnarray}
\braket{\Phi_{\alpha}}{\Phi_{\alpha}} &= Z_{cl} = \sum_c e^{-\beta_{cl} E_{cl} \left( C \right)} \\
E_{cl} \left( C \right) &= -u N_f \left( C \right),\quad \alpha = \exp\left(-\beta_{cl} u \over 2 \right) \label{intCDM}.
\end{eqnarray}
All equal time correlation functions of $\ket{\Phi_{\alpha}}$ are equal those of the classical model. This relationship allows us to write down a local RK-like Hamiltonian \cite{Ardonne2004,Castelnovo2005a} for which $\ket{\Phi_{\alpha}}$ is an exact zero energy ground state:
\begin{eqnarray}
\fl H_\alpha \equiv   \sum_{p} - t\Bigl(\ketbra{\plaqa}{\plaqb}+h.c.\Bigr)+ t\Bigl( \alpha^{- \Delta N_f / 2} \ketbra{\plaqa}{\plaqa} &+ \alpha^{ \Delta N_f / 2}\ketbra{\plaqb}{\plaqb} \Bigr), \label{Halpha} 
\end{eqnarray}
with $\Delta N_f \equiv N_f(\plaqb) - N_f(\plaqa)$ and $N_f(\plaqb)$ ($N_f(\plaqa)$) is the total number of flippable plaquettes on the lattice with $p$ in orientation $\plaqb$ ($\plaqa$). Here we see that $\alpha =1 \Rightarrow u=0$, which is the non-interacting point of the classical model; consequently $H_{\alpha = 1} = \Hrk$ at the RK point ($v=t$). The interacting classical dimer model has previously been studied using transfer matrix techniques by Trousselet {\it et al.}  \cite{Trousselet2007}. For $u>0$ ($\alpha < 1$), $E_{cl}$ is minimized by configurations which have the maximum number of flippable plaquettes, $N_l/6$. There is a large number of such configurations, including the 12 symmetry related columnar states and all configurations related to these by translating rows of dimers (row shifting modes) or rotating all dimers in a set of columns (column shifting modes). For $\alpha = 0$, this degeneracy prevents the formation of local order. However an ''order-by-disorder'' mechanism~\cite{Villain1980a} might allow for fluctuations to favor an ordered state for nonzero $\alpha$. Indeed the results of Ref. \cite{Trousselet2007} are consistent with a first order phase transition from a liquid phase for $\alpha>\alpha_c$ to an ordered phase for $\alpha<\alpha_c$, where the row shifting modes favor configurations with dimers aligned in the same direction.

Local updates do not give access to the global defects that appear as $\alpha \rightarrow 0$. Therefore we have implemented a directed loop Monte Carlo algorithm~\cite{Sandvik2006,Syljuasen2004}  to sample $\ket{\Phi_{\alpha}}$ on lattice sizes up to $L = 128$, using $\vert R_{col} \rangle$ as the reference configuration. The dimer model directed-loop algorithm generates non-local updates by creating a pair of defects that violate the hard core, fully packed dimer constraint; these defects undergo a directed random walk around the lattice until they coincide and annihilate~\cite{Sandvik2006}. The path of these defects forms a closed loop, along which the dimer configuration is updated. To sample $\vert \Phi_{\alpha} \rangle$, we have chosen the local weights of each step in the random walk to reflect $\vert \Phi_{\alpha} \rangle$~\cite{Syljuasen2002}, and found a solution to the {\em directed-loop equations}~\cite{Syljuasen2004} (see \ref{App} for details). This algorithm is ergodic in all topological sectors, and therefore the results presented here are computed over all winding sectors. 

\begin{figure}[]
\begin{centering}
$\begin{array}{c c}
  \includegraphics[width=3in]{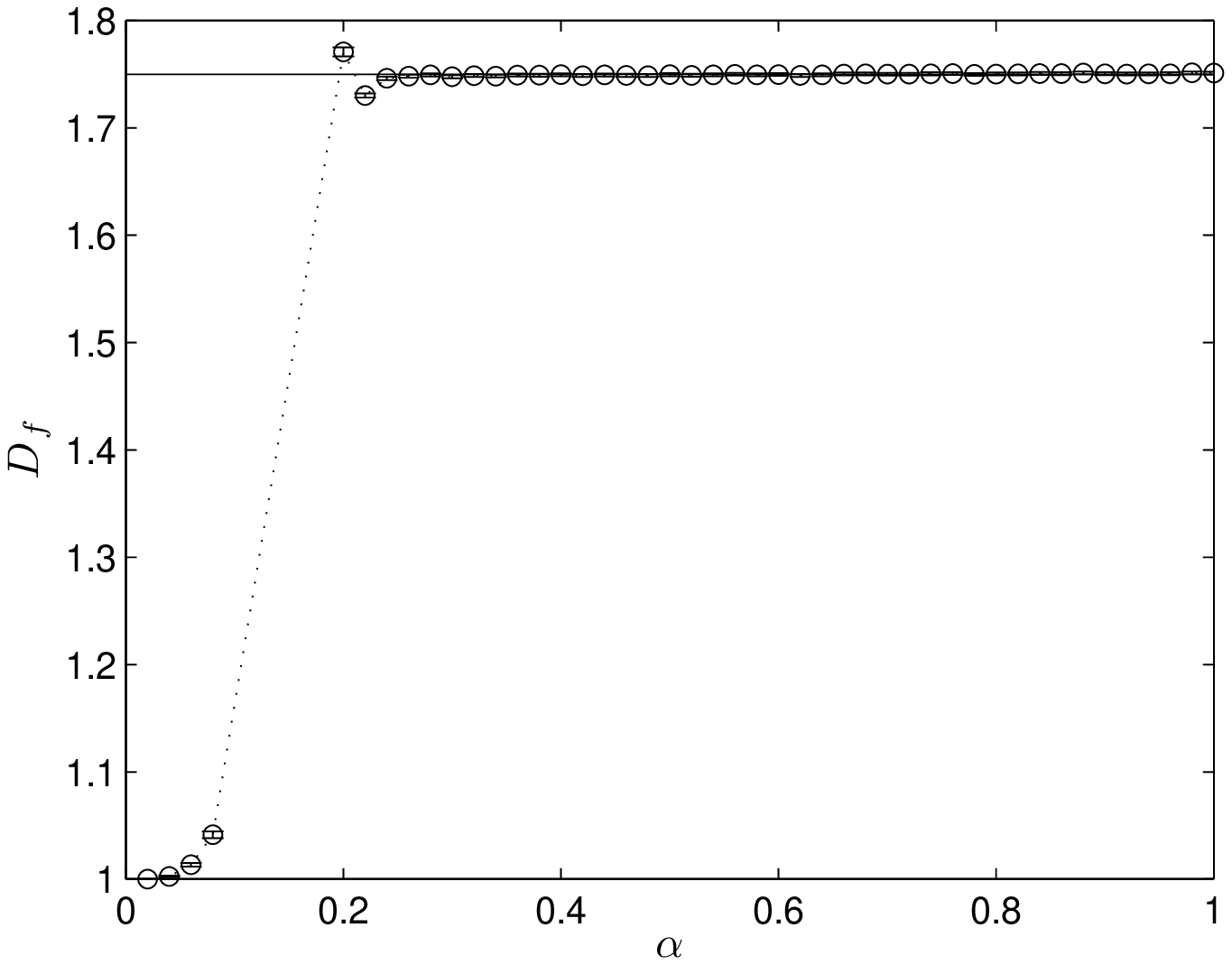} &
  \includegraphics[width=3in]{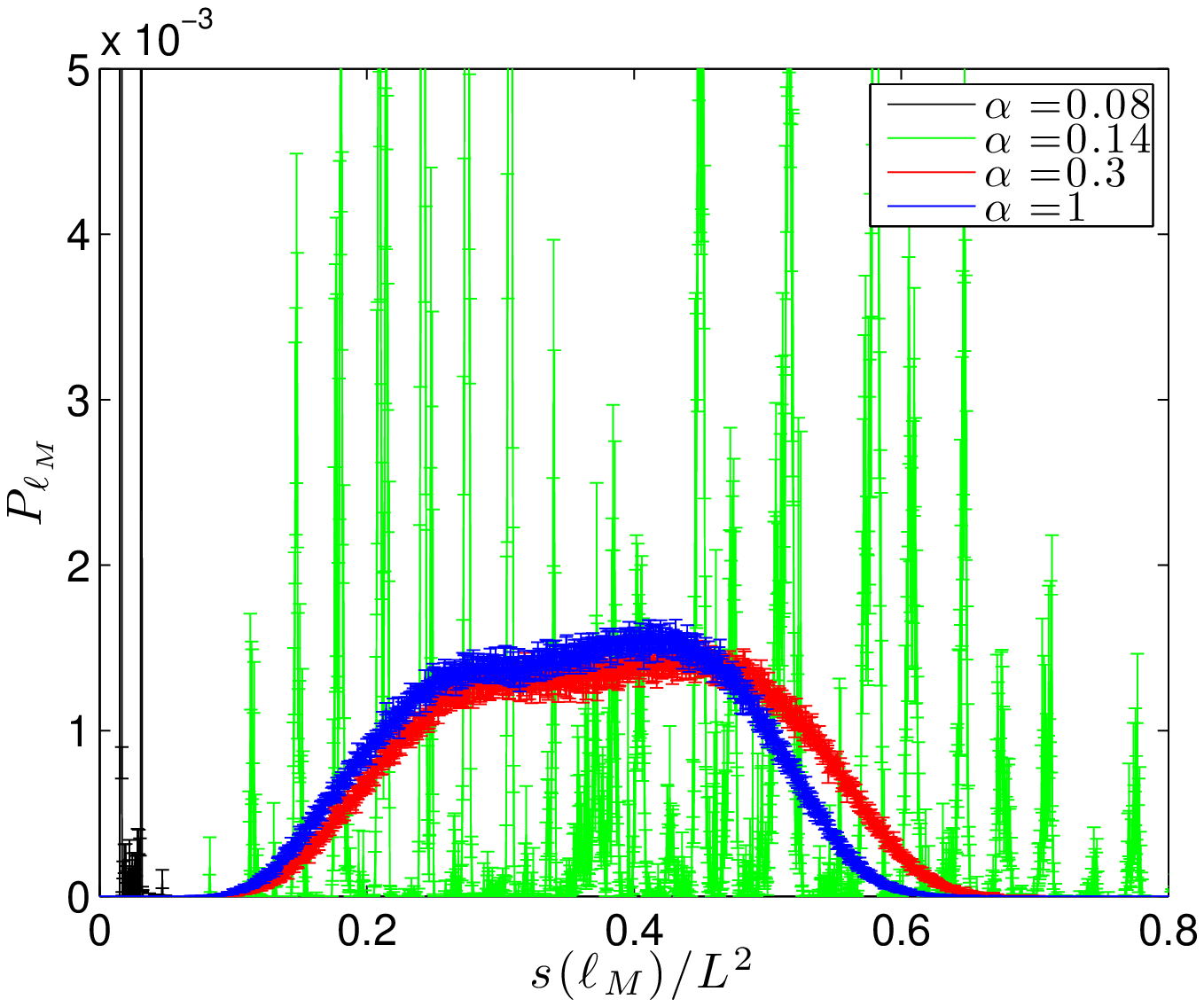} \\
  \mbox{\bf (a)} & \mbox{\bf (b)}
\end{array}$
\end{centering}
\caption{
(a) Fractal dimension of the largest loop $\ell_M$ of $\ket{\Phi_{\alpha}}$ computed with a directed loop algorithm for lattice sizes up to $L = 128$. The solid line corresponds to  $D_f = 7/4$, the value at the RK point. The dotted line is a guide for the eye; the gap in data is where the finite size scaling of $s(\ell_M)$ doesn't converge to a power law for the lattice sizes studied. (b) Longest loop distribution of $\ket{\Phi_{\alpha}}$ on a $L = 64$ lattice, for several values of $\alpha$, computed using $\vert R_{col} \rangle$. For very small $\alpha$, $P_{\ell_M}$ is peaked at $L$, consistent with an ordered phase with all dimers aligned in the same direction.
}\label{fignfplp}
\end{figure}

Figure \ref{fignfplp} shows that in the presence of interactions, the dimer liquid persists down to $\alpha \sim 0.2$. For much smaller values of $\alpha$, $D_f$ approaches $1$, indicative of a symmetry broken phase. However there is an intermediate regime where the finite size scaling of $\ell_M$ does not converge to a power law (this is shown by the dotted line in figure \ref{fignfplp}). The longest loop distrbution $P_{\ell_M}$ (right panel of figure \ref{fignfplp}) displays a series of peaks in this intermediate regime, whereas for small $\alpha$, there is only a single peak at $s(\ell_M) = L$. The small $\alpha$ regime is consistent with the low temperature phase of Ref. \cite{Trousselet2007} where row shifting modes dominate; with this choice of reference dimerization row defects are loops of length $L$ winding around the lattice in the direction of the rows. 

To further characterize the appearance of this symmetry breaking order for small $\alpha$, we define a rotational symmetry breaking order parameter 
\begin{eqnarray}
M_{\mathrm{rot}}^2 \equiv \frac{1}{2 N_d^2} \sum_{i,j} \left( N_d^i - N_d^j \right) ^2,
\end{eqnarray}
where $i$ and $j$ each represent one of the three possible dimer orientations on the triangular lattice, and $N_d^i$ is the number of dimers in a given orientation. Additionally, we define a columnar order parameter, 
\begin{eqnarray}
M_{\mathrm{col}}^2 \equiv \frac{1}{N_d^2} \sum_{(r,c)} \left( N_d \left[ R_{col}^{(r,c)} \right]  - N_d \left[ \tilde{R}_{col}^{(r,c)} \right] \right) ^2,
\end{eqnarray}
where $R_{col}^{(r,c)}$ is a columnar configuration with rows oriented in direction $r$ and columns in direction $c$, and $\tilde{R}_{col}^{(r,c)}$ is obtained by translating all dimers by one lattice spacing in direction $r$. $N_d \left[ R \right]$ is the number of dimers that coincide with the configuration $R$. In a columnar ordered phase, the expectation values of both $M_{\mathrm{rot}}^2$ and $M_{\mathrm{col}}^2$ saturate to $1$, whereas $M_{\mathrm{rot}}^2 = 1$ and $M_{\mathrm{col}}^2 = 0$ in the rotational symmetry broken ordered phase that is favored by the row shifting modes. 

\begin{figure}[h] 
   \centering
  \includegraphics[width=6in]{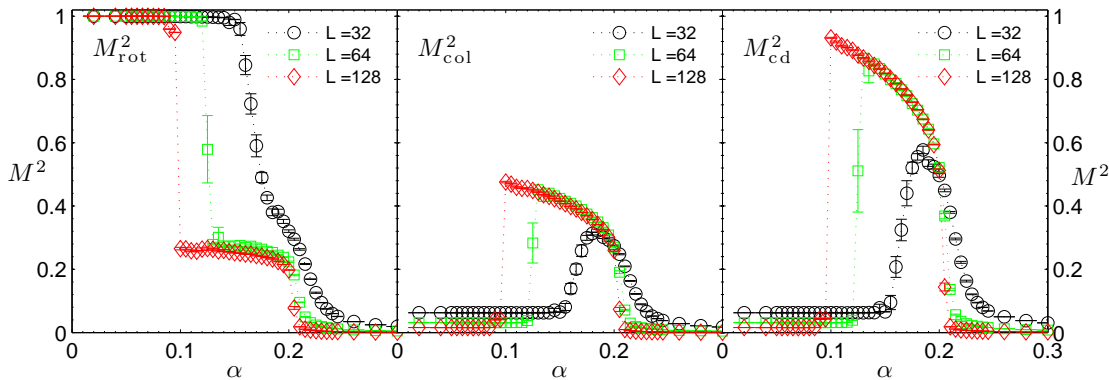} 
 \caption{ 
Expectation values of dimer order parameters $M_{\mathrm{rot}}^2$ (left), $M_{\mathrm{col}}^2$(center), $M_{\mathrm{cd}}^2$(right) in $\ket{\Phi_{\alpha}}$ in the symmetry broken phases. For larger values of $\alpha$, all local order parameters vanish.
}\label{figM2}
\end{figure}

Figure \ref{figM2} shows the expectation values of these dimer order parameters in the symmetry broken phases. We see that for small $\alpha$, $M_{\mathrm{rot}}^2$ saturates to $1$, whereas $M_{\mathrm{col}}^2$ scales to zero for larger system sizes (figure \ref{figM2}). This corresponds to the rotational symmetry broken phase where row defects dominate and destroy the columnar order, as discussed in Ref. \cite{Trousselet2007}. However, there is an intermediate regime for larger system sizes where the expectation value of $M_{\mathrm{rot}}^2$ reaches at plateau at approximately $1/4$, and $M_{\mathrm{col}}^2$ is finite and approaching $1/2$. This is consistent with a phase for which column shifting modes dominate, such that two dimer directions are preferred. The ideal columnar defect configurations saturate the order parameter
\begin{eqnarray}
\fl M_{\mathrm{cd}}^2 \equiv \frac{1}{N_d^2} \sum_{(r,r',c)} \left( N_d \left[ R_{col}^{(r,c)} \right] +N_d \left[ R_{col}^{(r',c)} \right]  - N_d \left[ \tilde{R}_{col}^{(r,c)} \right] -N_d \left[ \tilde{R}_{col}^{(r',c)}\right]\right) ^2\label{Mcd}
\end{eqnarray}
to $1$, while $M_{\mathrm{rot}}^2$ and $M_{\mathrm{col}}^2$ are $1/4$ and $1/2$, respectively. In (\ref{Mcd}), the sum is over $r \neq r' \neq c$. Figure \ref{figM2} shows that, in the intermediate phase, $M_{\mathrm{cd}}^2$ approaches 1 with decreasing $\alpha$. We note that while perfect column defects cost zero (classical) energy, adding a horizontal kink to the column defect costs a finite energy; the increased degeneracy of kinked defects may favor them over perfect defects for sufficiently large $\alpha$. If kinked column defects proliferate, they will generate a finite (but non-maximal) expectation value of $M^2_{cd}$. This suggests that for $\alpha \rightarrow 0$, global row defects dominate but there is an intermediate regime where kinked column defects dominate. Both of these transitions appear to be first order as shown by the discontinuities seen in these order parameters for larger system sizes (see figure \ref{figM2}); we have confirmed this by observing a double peak structure in the histograms of these order parameters at the transitions. Finite size scaling of these transitions suggests that both transitions occur at finite $\alpha$ in the thermodynamic limit.

 The distinction between these results and what is seen in Ref. \cite{Trousselet2007} is likely due to the difference in the geometry of the systems studied. We have also studied triangular lattices with $L_x \neq L_y$, where $L_{x,y}$ are the lengths in the two lattice directions. We find that as $L_y/L_x$ increases, the width of the intermediate regime decreases. This follows from the discussion of Ref. \cite{Trousselet2007}: increasing $L_y/L_x$ increases the number of row-shifting modes, and therefore favors the row defect ordering. Ref. \cite{Trousselet2007} considers lattices with $L_y \gg L_x$, so the results of that work are consistent with ours in this limit.
  
\section{d-isotopic quantum loop gas in the triangular lattice QDM} 
\label{sec:diso}

\begin{figure}[] 
   \centering
  \includegraphics[width=2in]{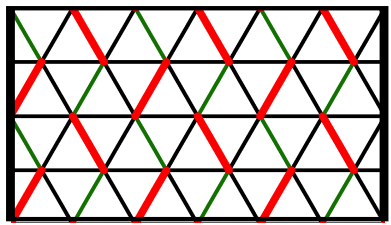} 
 \caption{ The decorated triangular lattice from reference \cite{Freedman2005a} used to define $\ket{\Psi_d}$. Here the staggered configuration defined by the red links is used as the reference dimerization.
}\label{fig_dqlg_lat}
\end{figure}

So far we have only considered transitions out of the $O(1)$ dimer loop liquid to crystalline phases; we now consider a wavefunction that allows us to explore other loop gases in the triangular lattice QDM. In reference \cite{Freedman2005a} Freedman \textit{et. al.} introduced a generalization of the quantum loop gas of the toric code they termed the $d$-isotopic loop gas \cite{Freedman2005a}. By giving a weight $d$ to each contractible loop in the ground state wavefunction, the $O(1)$ loop liquid of the toric code at $d=1$ can be driven to loop gas phases for $O(n>1)$:
\begin{equation}
\ket{\Psi_d} = \sum_{C} d^{\mathcal{N}_c } \ket{C} \label{psi_d}.
\end{equation}
In equation \eref{psi_d}, $\mathcal{N}_c$ is the number of contractible loops in $\ket{C}$ and the sum over $C$ is taken over all configurations in the same winding sector. The norm of $\Psi_d$ is equal to classical partition function of the $O(n)$ loop model for $n=d^2$~\cite{Freedman2005b}. Freedman \textit{et. al.} also wrote down an RK-like Hamiltonian, $H_d$ for which $\ket{\Psi_d}$ is the exact zero energy ground state \cite{Freedman2005a}. Every winding sector contains a unique ground state; as such this d-isotopic loop gas has a large degeneracy. For $1\leq d\leq \sqrt{2}$, $\ket{\Psi_d}$ is a critical loop liquid and it  has been shown that $H_d$ is gapless. For $n > 2$, the $O\left(n\right)$ model is in a short looped phase~\cite{Nienhuis1984}, and correspondingly, the d-isotopic loop gas is in a short looped phase for $d > \sqrt{2}$. All diagonal local correlation functions in $\ket{\Psi_d}$ vanish exponentially; however Troyer et. al. showed certain that off-diagonal correlation functions have a power-law decay \cite{Troyer2008}. The original motivation for introducing $\ket{\Psi_d}$ was the search for gapped topological phases beyond the abelian toric code phase. Subsequent work \cite{Troyer2008} shows that adding local interactions to $H_d$ to open a gap drives the loop gas to the toric code phase. Consistent with the disordered nature of the liquid phase of $\ket{\Psi_d}$, Troyer et. al. showed that the critical phase of the loop liquid could be described by the fractal dimension of the long fractal loop \cite{Troyer2008}.

To implement $\Psi_d$ in a quantum dimer model we color the triangular lattice according to reference \cite{Freedman2005a}. Links occupied in a staggered reference configuration $\vert R_0 \rangle$ are colored red and links connecting the red links are colored green (see figure \ref{fig_dqlg_lat} ). Here we choose a slightly different interpretation of the transition loop graph; dimers in $\ket{C}$ that coincide with $\vert R_0 \rangle$ are considered to be minimal, length 2 loops. As such, the loop coverings defined by the transition graphs are fully packed loop coverings. In this case, $\Psi_d$ is described by a fully packed loop model, which in turn is related to the dense phase of the $O(n)$ model~\cite{Blote1994}. Consequently, the geometric exponents as defined in section \ref{sec:RK} are the same in the fully packed loop model as those of the dense phase of the $O(n)$ model with $d^2 = n$~\cite{Saleur1987,Kondev1996}.

With this coloring, a local quantum dimer model Hamiltonian, $H_d$ can be defined such that $\ket{\Psi_d}$ defines the unique, zero energy ground state in each winding sector. To allow for the transition to the short looped, staggered crystal phase in the limit $d \rightarrow \infty$, 4-dimer dynamics that flip dimers around a length 8 loop are included in $H_d$ to connect the staggered configurations to other configurations. Additionally, following reference \cite{Freedman2005a}, we wish to define $H_d$ such that it  only connects configurations in the same loop winding sector, not in a parity defined winding sector. This requires removing the plaquette flip terms that can act as a surgery and connect different winding sectors--this is done by excluding plaquette flips in $H_d$ for rhombi that have a green diagonal link. To compensate for a lost ergodicity due to the removal of these plaquette flip terms, a 3-dimer resonance around a length 6 triangular loops is also added to $H_d$~\cite{Freedman2005a}.

\begin{figure}[]
\begin{center}
$\begin{array}{c c}
  \includegraphics[width=3in]{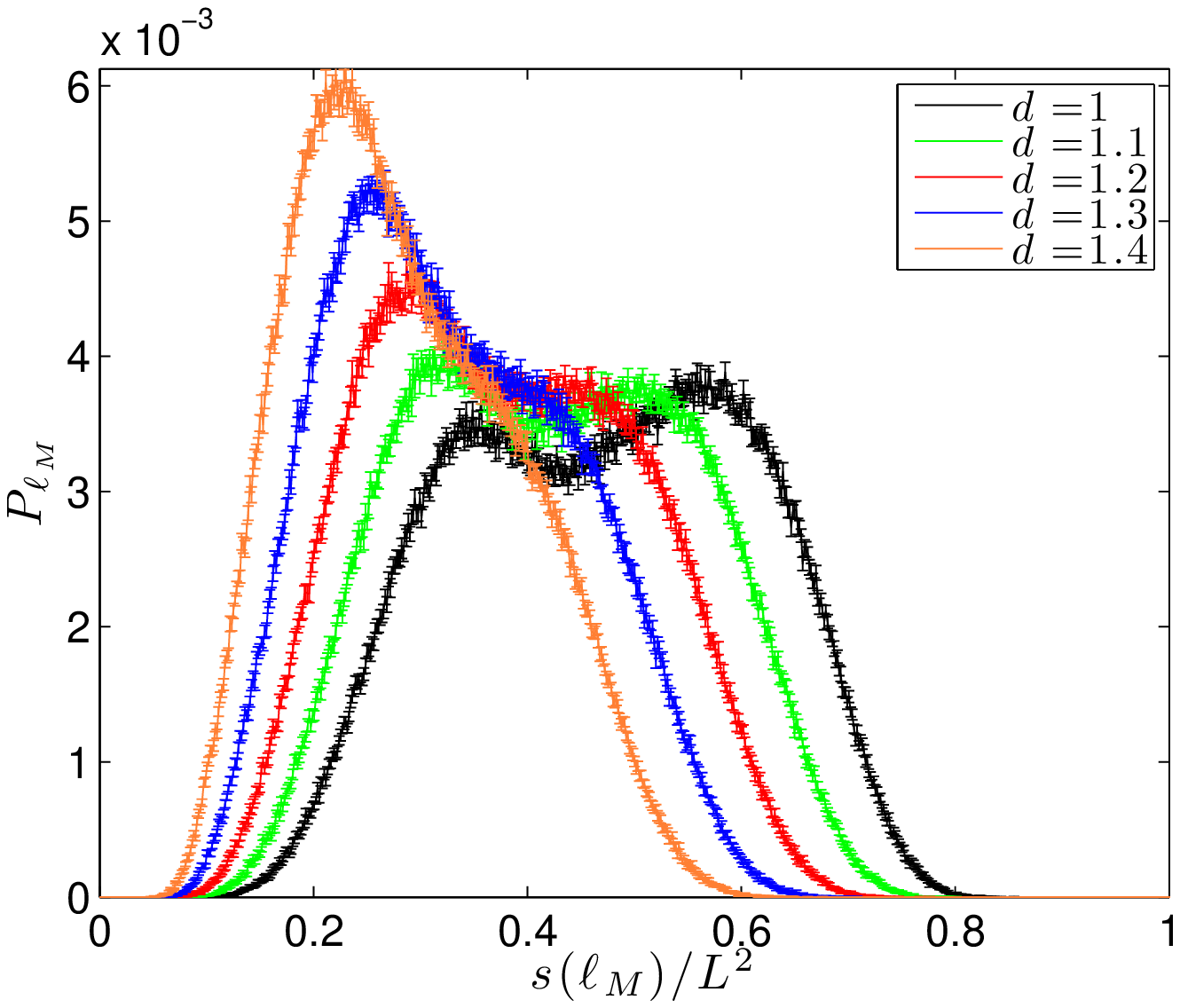} &
  \includegraphics[width=3in]{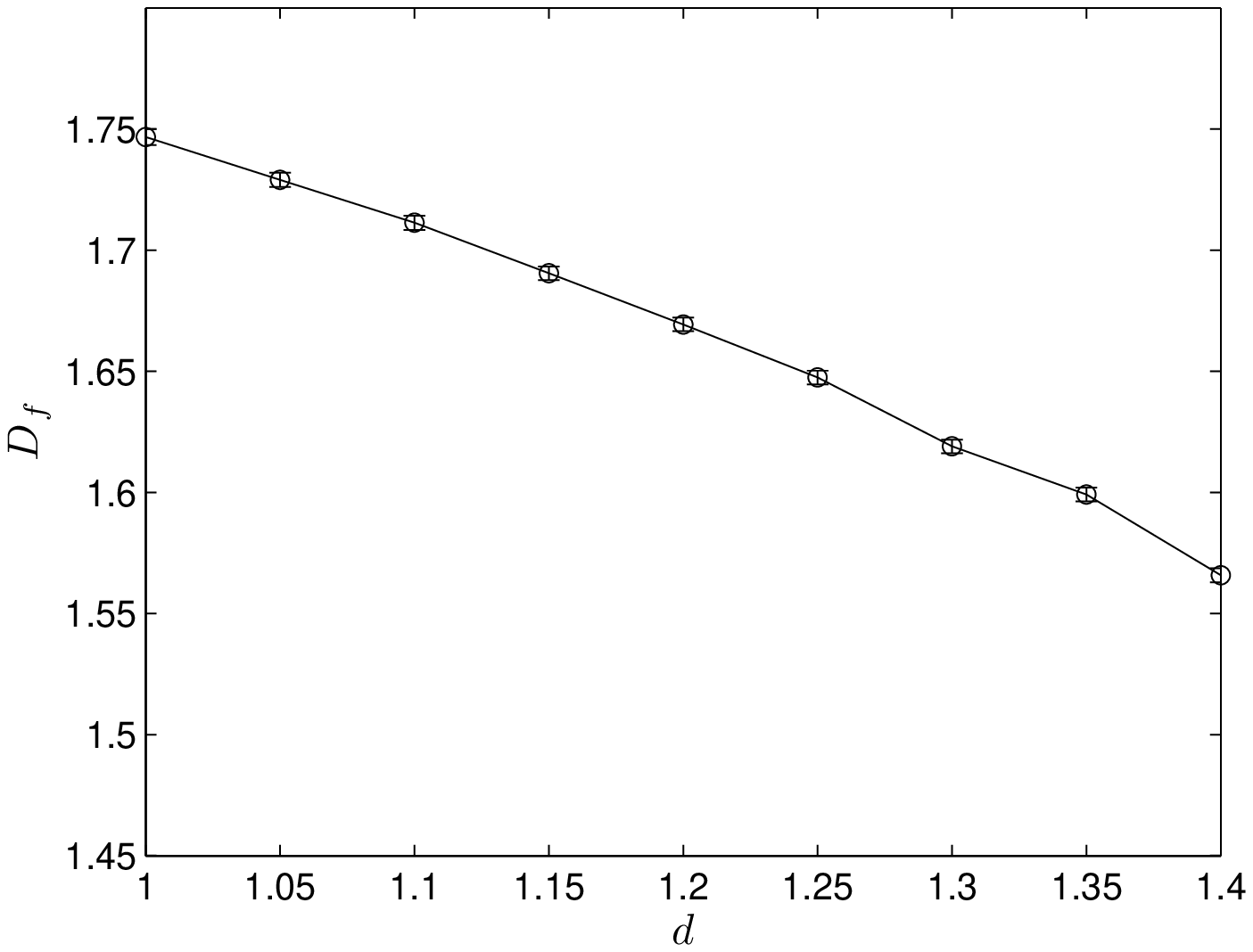} \\
  \mbox{\bf (a)} & \mbox{\bf (b)}
\end{array}$
\end{center}
\caption{(a) Longest loop distribution $P_{\ell_M}$ for $\ket{\Psi_d}$ in the zero winding sector on a $L=24$ lattice  for several values of $d$. (b) Fractal loop dimension of $\ket{\Psi_d}$, computed from finite size scaling of $s(\ell_M)$ on lattice sizes up to $L = 84$.
}\label{fig_dqlg_lM}
\end{figure}

Figure \ref{fig_dqlg_lM}a shows the longest loop distribution for several values of $d$ computed by Monte Carlo sampling of $\ket{\Psi_d}$ in the triangular lattice quantum dimer model. We see that a liquid phase persists for $d > 1$. Figure \ref{fig_dqlg_lM}a shows that the fractal dimension is a continuously varying function of $d$. Thus $\ket{\Phi_d}$ can describe dimer liquid phases distinct from the $Z_2$ topologically ordered phase at the RK point.

\section{Off-diagonal loop operators in the triangular lattice QDM}
\label{sec:wilson}

\begin{figure}[] 
\centering
  \includegraphics[width=3in]{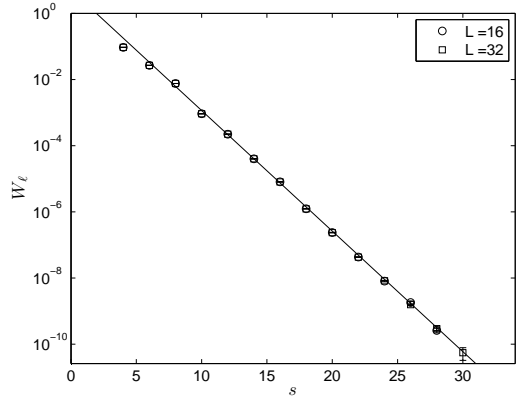} 
 \caption{ 
Expectation value of off-diagonal loop operator $W_\ell$ as a function of loop lenth $s$ at the RK point, computed for various parallelograms of different lengths on lattices of size $L = 16,32$. The line shows a fit to an exponential decay, indicating a clear perimeter law behavior.
}
\label{fig_WilsonL12RK}
\end{figure}

Off-diagonal operators in a quantum dimer model that do not violate the hard-core constraint are loop operators: two dimer configurations can be connected by flipping dimers along their transition loops, as given in equation \eref{Wtrans}. In the toric code, such loop operators are related to the Wilson loop of the underlying $Z_2$ gauge theory. In, the ground state $\vert \Psi_{\mathrm{TC}} \rangle$, the expectation value of all loop operators is exactly 1 (a "zero law"). However the addition of a magnetic field will cause the loop operators to decay with the length of the loop (a "perimeter law"). In a quantum dimer model, the fully-packed constraint means that $\left< W_\ell (s) \right> < 1$ for all loops, even at the RK point~\cite{Misguich2008d}. Moessner, Sondhi and Fradkin suggested that $\left< W_\ell (s) \right>$ will decay as a perimeter law at the RK point due to the extensive entropy of dimerizations of the lattice~\cite{Moessner2001}. Figure \ref{fig_WilsonL12RK} shows $\expect{W_\ell}_{\mathrm{RK}}$ computed for a set of parallelograms of differing lengths on the triangular lattice by Monte Carlo sampling of $\vert RK \rangle$. $\langle W_\ell (s) \rangle$ is seen to display a clear perimeter law at the RK point.

\section{Conclusions} 
In this paper we have characterized the formation of topological order in quantum dimer models in terms of the physical mechanism of loop condensation. This allows for a more direct connection between the $Z_2$ topologically ordered liquid of the triangular lattice QDM to that of the toric code and other lattice models. We have shown that geometric properties of the transition graph loop distribution may be used to distinguish dimer liquid and crystal phases. Additionally, the fractal dimension of the spanning loop may be used to distinguish dimer liquid phases. We believe this approach may be useful for the study of dimer liquids and the full phase diagram of the canonical quantum dimer model Hamiltonian.

\ack
This research was supported by NSF Grant No. PH4-0803429. CMH thanks Jonathan Dubois for help in the preparation of Monte Carlo simulations.

\appendix
\section{The directed-loop algorithm in the interacting dimer model}
\label{App}
The directed-loop algorithm~\cite{Syljuasen2002,Sandvik2006} generates non-local updates in a dimer model by creating a pair of defects (a doubly occupied vertex and an unoccupied vertex) that violate the fully-packed, hard-core dimer constraint. One of these defects undergoes a directed random walk and, when the two defects coincide once again, this generates a new allowable dimerization, where the dimers have been flipped along the path of the path of the defect~\cite{Sandvik2006,Syljuasen2004}. To maintain detailed balance, the probabilities of choosing each direction on a given step of the random walk must satisfy the directed-loop equations~\cite{Syljuasen2002}. 

Consider a step in the loop-update where the loop has entered a vertex $v$ from link $l$. We define the probability of choosing link $l'$ to exit the vertex as $P_{l,l'} \equiv t_{l,l'}/w_l$, where $w_l$ is the weight of the configuration with the dimer on link $l$. Detailed balance and probability conservation require
\begin{eqnarray}
\sum_{l'} t_{l,l'} = w_l,\quad\quad t_{l,l'} = t_{l',l}
\end{eqnarray}
For a given set of weights $\{w_l\}$, this leads to a set of directed-loop equations.

Here we consider sampling the square of the wavefuction $\Phi_\alpha (l) = \alpha ^{-N_{fp}}$. There are three possible weights for a dimer to occupy a link $\{w_0,w_1,w_2\}$, corresponding to a link with $0$, $1$ or $2$ parallel dimers. The choice $w_2 = 1$, so $w_2 = w_1^2 = \alpha^4$ gives the following directed-loop equations:
\begin{eqnarray}
1 &= \left(6-n_1-n_2 \right) t_{02} + n_1 t_{12}+\left(n_2-1\right)t_{22} + t_2^b \label{dl1}\\
\alpha^2 &= \left(6-n_1-n_2 \right)t_{01} + n_2 t_{12}+\left(n_1-1\right)t_{11}+t_1^b \label{dl2}\\
\alpha^4 &= n_1t_{01} + n_2 t_{02} + \left(6-n_1-n_2\right) t_{00} + t_0^b. \label{dl3}
\end{eqnarray}
In (\ref{dl1}-\ref{dl3}), $t_{pq}$ is the weight for a transition between links with $p$ and $q$ parallel dimers, $n_p$ is the number of links with $p$ parallel dimers at vertex $v$, and $t_p^b$ is rate to "bounce" back from a link with $p$ parallel dimers. Since these equations are underdetermined, we can choose a solution to (\ref{dl1}-\ref{dl3}) for which the bounce probabilities vanish at the RK point ($\alpha = 1$):
\begin{eqnarray}
t_{00} &= t_{01} = t_{02} = \frac{\alpha^4}{5}, \quad t_{12} = \frac{\alpha^2}{5}, \quad t_0^b = 0, \\
 t_{11} &= \left\{ \begin{array}{lr} \frac{\alpha^2 \Bigl( \alpha^2 n_1 - \left(1 - \alpha^2 \right) n_2 + 6\left(1- \alpha^2 \right) -1 \Bigr) }{5\left(n_1-1\right)} & n_1 > 1 \\
0 & n_1 \leq 1\end{array} \right., \\
t_{22} &= \left\{ \begin{array}{lr}\frac{ \alpha^4 n_2 - \alpha^2 \left(1 - \alpha^2 \right) n_1 + 6\left(1- \alpha^4 \right) -1 }{5\left(n_2-1\right)}& n_2 > 1 \\
0 & n_2 \leq 1 \end{array} \right., \\
 t_1^b  &=  \left\{ \begin{array}{lr} \frac{\alpha^2 \left( 1 - \alpha^2 \right) \left( 5 - n_2 \right)}{5} & n_1 = 1 \\
0 & n_1 \neq 1\end{array} \right. ,\\
t_2^b  &=  \left\{ \begin{array}{lr} 1- \frac{\alpha^2}{5} \Bigl( 5\alpha^2+(1- \alpha^2 )n_1 \Bigr) & n_2 = 1 \\
0 & n_2 \neq 1\end{array} \right. .
\end{eqnarray}
The directed loop algorithm using this solution to (\ref{dl1}-\ref{dl3}) has be compared to a local update algorithm and is sufficiently efficient in the regimes studied.
\section*{References}
\bibliography{LCinTQDM}

\end{document}